\documentclass[letterpaper,12pt]{article}

\usepackage{mathptmx}

\usepackage{graphics}

\usepackage[authoryear]{natbib} 
\usepackage{url}

\usepackage{multirow}
\usepackage{graphicx}
\usepackage{bbm}
\usepackage{amssymb}
\usepackage{amsmath}
\usepackage{amsfonts}
\usepackage{amsthm}
\usepackage{rotating}
\usepackage{xr}
\usepackage{subfig}
\usepackage{setspace}
\usepackage{fullpage}

\def \s2{\sigma^2}
\def \hs2{\hat{\sigma}^2}
\def \siga2{\sigma_{\alpha}^2}
\def \sige2{\sigma_{\epsilon}^2}
\def \sig2{\sigma^2}

\newcommand{\beq}{\begin{equation}}       
\newcommand{\eeq}{\end{equation}}       
\newcommand{\beqn}{\begin{eqnarray}}
\newcommand{\eeqn}{\end{eqnarray}}

\newcommand{\indicatorBig}[1]{\mathbbm{1}{\left[ {#1} \right] }}

\begin{document}

\begin{titlepage}

\begin{center}

{\LARGE \bf Who's good this year? Comparing the Information Content of Games in the Four Major US Sports} \\

\ \\

{\large Julian Wolfson and Joseph S. Koopmeiners \\
\ \\
Division of Biostatistics, University of Minnesota, Minneapolis, Minnesota }\\
\ \\
{\large \it Corresponding author's email address: julianw@umn.edu} \\
\ \\
{\large \today}
\end{center}

\begin{abstract}
In the four major North American professional sports (baseball, basketball, football, and hockey), the primary purpose of the regular season is to determine which teams most deserve to advance to the playoffs. Interestingly, while the ultimate goal of identifying the best teams is the same, the number of regular season games played differs dramatically between the sports, ranging from 16 (football) to 82 (basketball and hockey) to 162 (baseball). Though length of season is partially determined by many factors including travel logistics, rest requirements, playoff structure and television contracts, it is hard to reconcile the 10-fold difference in the number of games between, for example, the NFL and MLB unless football games are somehow more ``informative'' than baseball games. In this paper, we aim to quantify the amount of information games yield about the relative strength of the teams involved. Our strategy is to assess how well simple paired comparison models fitted from $X$\% of the games within a season predict the outcomes of the remaining ($100-X$)\% of games, for multiple values of $X$. We compare the resulting predictive accuracy curves between seasons within the same sport and across all four sports, and find dramatic differences in the amount of information yielded by individual game results in the four major U.S. sports.
\end{abstract}

\end{titlepage}

\newpage
\section{Introduction}

In week 14 of the 2012 NFL season, the 9-3 New England Patriots squared off on Monday Night Football against the 11-1 Houston Texans in a game with major implications for both teams. At the time, the Texans had the best record in the AFC and were in line to earn home-field advantage throughout the playoffs, while the New England Patriots had the best record in their division and were hoping to solidify their playoff position and establish themselves as the favorites in the AFC. The Patriots ultimately defeated the Texans 42-14, which led some commentators to conclude that the Patriots were the favorites to win the Super Bowl \citep{Walker2012,MacMullan2012} and that Tom Brady was the favorite for the MVP award \citep{Reiss2012}, while others opined that the Texans were closer to pretenders than the contenders they appeared to be for the first 13 weeks of the season \citep{Kuharsky2012}. These are strong conclusions to reach based on the results of a single game, but the power of such ``statement games'' is accepted wisdom in the NFL. In contrast, it is rare for the outcome of a single regular-season game to create or change the narrative about a team in the NBA, NHL, or MLB. While one might argue that the shorter NFL season simply drives commentators to imbue each game with greater metaphysical meaning, an alternative explanation is that the outcome of a single NFL contest actually does carry more information about the relative strengths of the teams involved than a single game result in the other major North American professional sports. In this paper, we ask and attempt to answer the basic question: how much does the outcome of a single game tell us about the relative strength of the two teams involved?

In the four major North American professional sports (baseball, basketball, football, and hockey), the primary purpose of the regular season is to determine which teams most deserve to advance to the playoffs. Interestingly, while the ultimate goal of identifying the best teams is the same, the number of regular season games played differs dramatically between the sports, ranging from 16 (football) to 82 (basketball and hockey) to 162 (baseball). Though length of season is partially determined by many factors including travel logistics, rest requirements, playoff structure and television contracts, it is hard to reconcile the 10-fold difference in the number of games in the NFL and MLB seasons unless games in the former are somehow more informative about team abilities than games in the latter. Indeed, while it would be near-heresy to determine playoff eligibility based on 16 games of an MLB season (even if each of the 16 games was against a different opponent), this number of games is considered adequate for the same purpose in the NFL.

There is a well-developed literature on the topic of competitive balance and parity in sports leagues \citep{Owen2010,Horowitz1997,Mizak2005,Lee2010,Hamlen2007,Cain2006,Larsen2006,Ben-Naim2006a,Kesenne2000,Vrooman1995,Koopmeiners2012}. However, most papers focus on quantifying the degree of team parity over consecutive years along with the effects of measures taken to increase or decrease it. In papers which compare multiple sports, season length is often viewed as a nuisance parameter to be adjusted for rather than a focus of inquiry. Little attention has been directed at the question of how information on relative team strength accrues over the course of a single season. 

In this paper, we aim to quantify the amount of information each games yields about the relative strength of the teams involved. We estimate team strength via paired-comparison \citep{Bradley1952} and margin-of-victory models which have been applied to ranking teams in a variety of sports \citep{McHale2011, Koehler1982, Sire2009, Martin1999}. The growth in information about the relative strength of teams is quantified by considering how these comparison models fitted from $X$\% of the games in a season predict the outcomes of the remaining ($100-X$)\% of games, for multiple values of $X$ (games are partitioned into training and test sets at random to reduce the impact of longitudinal trends over the course of a season). We begin by describing the data and analysis methods we used in Section 2. Section 3 presents results from recent seasons of the four major North American sports, and compares the ``information content'' of games across the four sports. In Section 4 we discuss the strengths and limitations of our analysis.

\section{Methods}
\subsection{Data}
We consider game results (home and away score) for the 2004-2012 seasons for the NFL, the 2003-2004 to 2012-2013 seasons of the NBA, the 2005-2006 to 2012-2013 seasons of the NHL, and the 2006-2012 seasons of MLB. Game results for the NFL, NBA and NHL were downloaded from Sports-Reference.com \citep{pfr} and game results for MLB were downloaded from Retrosheet \citep{retrosheet}. Only regular season games were considered in our analysis. The NFL plays a 16-game regular season, the NBA and NHL play 82 regular season games and MLB plays 162 regular season games. 

\subsection{Methods}

Let $\mathcal{G}$ represent all the games within a single season of a particular sport. Our goal is to quantify the amount of information on relative team strength contained in the outcomes of a set of games $G \subset \mathcal{G}$, as the number of games contained in $G$ varies. We consider how well the results of games in the ``training set'' $G$ allow us to predict the outcomes of games in a ``test set'' $G' = \mathcal{G} \setminus G$. Specifically, given $G$ and $G'$, our information metric (which we formally define later) is the percentage of games in $G'$ which are correctly predicted using a paired comparison model applied to $G$.

We consider two types of paired comparison models in our work. Each game $g \in G$ provides information on the home team ($H_g = i$), away team ($A_g = j$) and the game result as viewed from the home team's perspective. When only the binary win/loss game result $W_g$ is considered, we fit a standard Bradley-Terry model \citep{Bradley1952, Agresti02},
\begin{equation}
logit \left(\pi_{i,j}\right) = \beta_{i} - \beta_{j} + \alpha,
\label{eq:BT}
\end{equation}
where $\pi_{i,j} = P(W_g = 1)$ is the probability that the home team, team $i$, defeats the visiting team, team $j$. $\beta_{i}$ and $\beta_{j}$ are the team strength parameters for teams $i$ and $j$, respectively, and $\alpha$ is a home-field advantage parameter.

We fit a similar model when the actual game scores are considered. In this context, home team margin of victory (MOV) $\Delta_g$ is recorded for each game; $\Delta_g$ is positive for a home team win, negative for a home team loss, and zero for a tie. The paired comparison model incorporating margin of victory is:
\begin{equation}
\mu_{i,j} = \delta_{i} - \delta_{j} + \lambda,
\label{eq:MOV}
\end{equation}
where $\mu_{i,j} = E(\Delta_g)$ is the expected margin of victory for the home team, team $i$, over the visiting team, team $j$. $\delta_{i}$ and $\delta_{j}$ are team strengths on the margin-of-victory scale for teams $i$ and $j$, respectively, and $\lambda$ is a home-field advantage on the margin-of-victory scale. 

Both models \eqref{eq:BT} and \eqref{eq:MOV} can be fit using standard statistical software, such as R \citep{rsoftware}. Given estimates $\hat \beta_i$, $\hat \beta_j$, and $\hat \alpha$ derived by fitting model \eqref{eq:BT} to a set of games $G$, a predicted home team win probability $\hat \pi_g$ can be derived for every game $g \in G'$ based on which teams $i$ and $j$ are involved. A binary win/loss prediction for the home team is obtained according to whether $\hat \pi_g$ is greater/less than 0.5. Given estimates $\hat \delta_i$, $\hat \delta_j$, and $\hat \lambda$ from fitting model \eqref{eq:MOV}, home team margin of victory $\hat \mu_g$ can similarly be predicted for every game in $g \in G'$. A binary win/loss prediction for the home team is obtained according to whether $\hat \mu_g$ is positive, negative, or zero. 

Our metrics for summarizing the amount of information on relative team strength available from a set of game results $G$ for predicting the outcomes of a set of games in $G'$ are simply the fraction of games that are correctly predicted by the paired comparison models:
\begin{align*}
\mathcal{I}^{BT}(G,G') &= \frac{ \sum_{g \in G'} W_g \indicatorBig{\hat \pi_g > 0.5}}{ | G' | } \ \ \text{for the Bradley-Terry model \eqref{eq:BT}}\\
\mathcal{I}^{MOV}(G,G') &= \frac{ \sum_{g \in G'} W_g \indicatorBig{\hat \mu_g > 0}}{ | G' | } \ \ \text{for the margin-of-victory model \eqref{eq:MOV}}
\end{align*}
where $\hat \pi_g$ and $\hat \mu_g$ are estimates derived from game results in $G$, and $|G'|$ denotes the number of games in $G'$. 

For a given season, training data sets $G_1, G_2, \dots, G_K$ were formed by randomly sampling games corresponding to X\% of that season. Test data sets $G'_1, G'_2, \dots, G'_K$ were created as the within-season complements of the training sets, i.e., if $G_k$ consists of a number of games corresponding to X\% of the season, then $G'_k$ contains the remaining (100-X)\% of games in that season. Training (and corresponding test) data sets were created for X\% = 12.5\%, 25.0\%, 37.5\%, 50.0\%, 62.5\%, 75.0\% and 87.5\% of the games in each available season. Games were sampled at random so as to reduce the influence of temporal trends over the course of a season, for example, baseball teams who are out of playoff contention trading away valuable players and giving playing time to minor league prospects in August and September. 

Information on relative team strength over a single season was computed and summarized as follows:

\begin{enumerate}
\item For X = 12.5, 25, 37.5, 50, 62.5, 75, and 87.5:
	\begin{enumerate}
	\item Generate 100 training sets $G_1, G_2, \dots, G_{100}$ (and complementary test sets $G'_1, G'_2, \dots, G'_{100}$) by randomly sampling X\% of games without replacement from $\mathcal{G}$.
	\item For each training set $G_k$:
		\begin{enumerate}
		\item Fit models \eqref{eq:BT} and \eqref{eq:MOV} to the games in $G_k$.
		\item Obtain binary win/loss predictions for the games in the test set $G'_k$.
		\item Evaluate the information metrics $\mathcal{I}^{BT}(G_k, G'_k)$ and $\mathcal{I}^{MOV}(G_k,G'_k)$
		\end{enumerate}
	\item Average the computed information metrics to estimate the predictive accuracy of paired comparison models fitted to data from X\% of the entire season ($\mathcal{I}^{BT}$ and $\mathcal{I}^{MOV}$).
	\end{enumerate}
\item Tabulate and plot $\mathcal{I}^{BT}$ and $\mathcal{I}^{MOV}$ across different values of X.
\end{enumerate}


The natural comparison value for our information metrics is the predictive accuracy of a naive model which chooses the home team to win every game. As shown in the plots below the average win probability for the home team (as determined by the parameters $\alpha$ and $\lambda$ in models \eqref{eq:BT} and \eqref{eq:MOV} respectively) varies from approximately 53\% to 61\% across the four sports we consider.  

\section{Results}

\subsection{National Football League}

\begin{figure}[!h]
\centering
\includegraphics*[width = 6in]{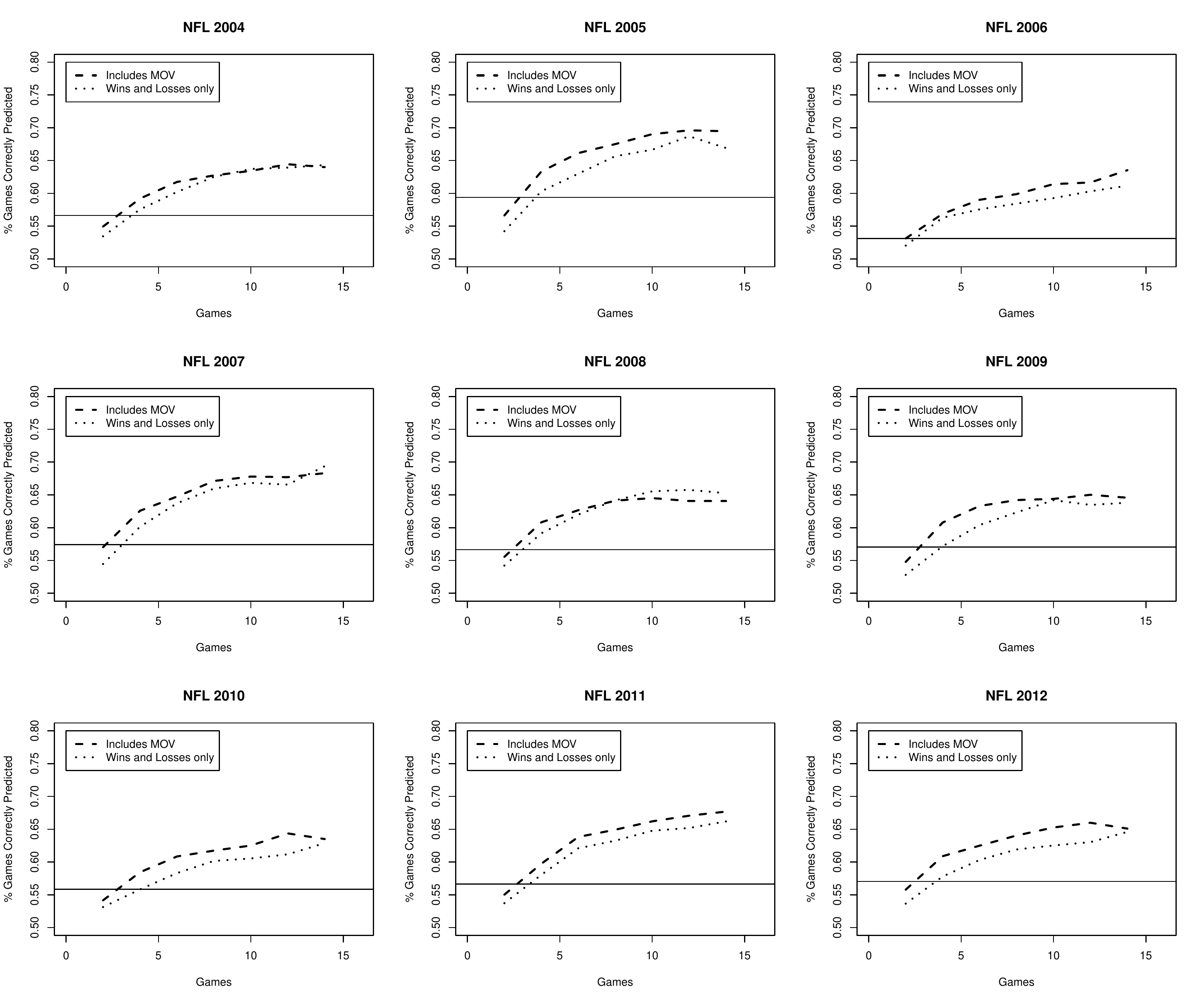}
\caption{Percent of games correctly predicted on test set vs. average number of games per team in training set, NFL seasons 2004-2012.\label{nfl_results}}
\end{figure}

Figure~\ref{nfl_results} plots the percent of games correctly predicted on the test set versus the average number of games per team in  the training set for the 2004-2012 National Football League seasons. Both paired comparison models (i.e., those which incorporate and ignore margin of victory) outperform simply picking the home team to win every game. The margin of victory model appears to perform slightly better than the paired comparison model, though the differences are modest and in some seasons (e.g., 2004 and 2008) are non-existent. The prediction accuracy of both models improves throughout the season in most seasons (years 2008 and 2009 being notable  exceptions), indicating that we are gaining information about the relative strengths of teams even in the final weeks of the season.

\subsection{National Basketball Association}
\begin{figure}[!h]
\centering
\includegraphics*[width = 6in]{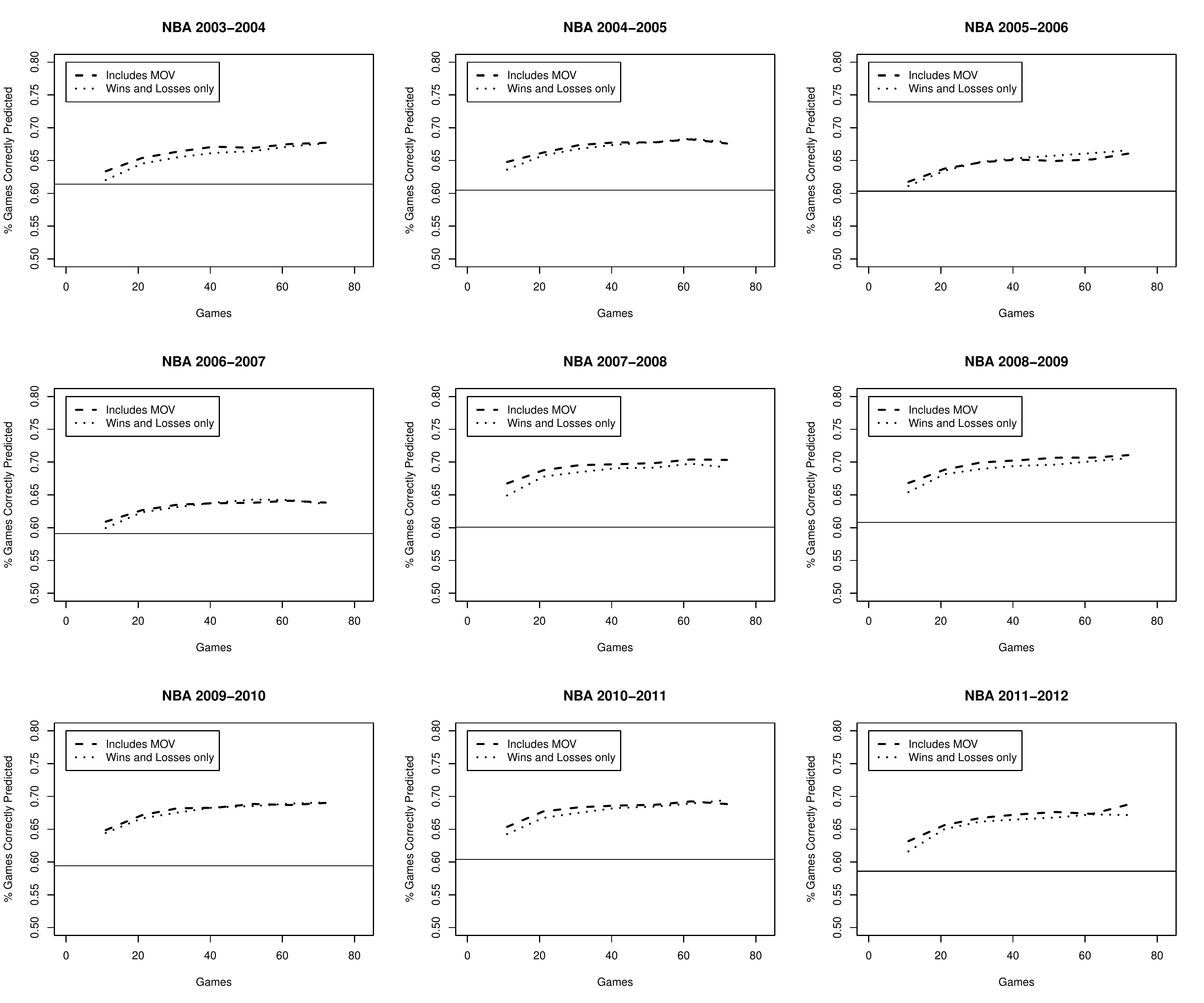}
\caption{Percent of games correctly predicted on test set vs. average number of games per team in training set, NBA seasons 2003-2013. \label{nba_results}}
\end{figure}

Results for the National Basketball Association can be found in Figure~\ref{nba_results}. The NBA was the most predictable of the four major North American professional sports leagues. Using 87.5\% of games as a training set, our model was able to accurately predict up to 70\% across seasons. The NBA also had the largest home court advantage with home teams winning approximately 60\% of games. There was virtually no advantage in including margin of victory in our model; indeed, it led to slightly worse predictions during the 05-06 season. The only major difference between the NFL and NBA was the growth of information over the season. While the accuracy of our predictions for the NFL continued to improve as more games were added to the training set, model accuracy for the NBA was no better when 75\% of games were included in the training set than when 25\% of games were included. Analyses using the \texttt{segmented} package in R for fitting piecewise linear models \citep{Muggeo2003, Muggeo2008} confirmed an inflection point in the prediction accuracy curve approximately 25-30 games into the season.

\subsection{Major League Baseball and the National Hockey League}

\begin{figure}[!h]
\centering
\includegraphics[width = 6in]{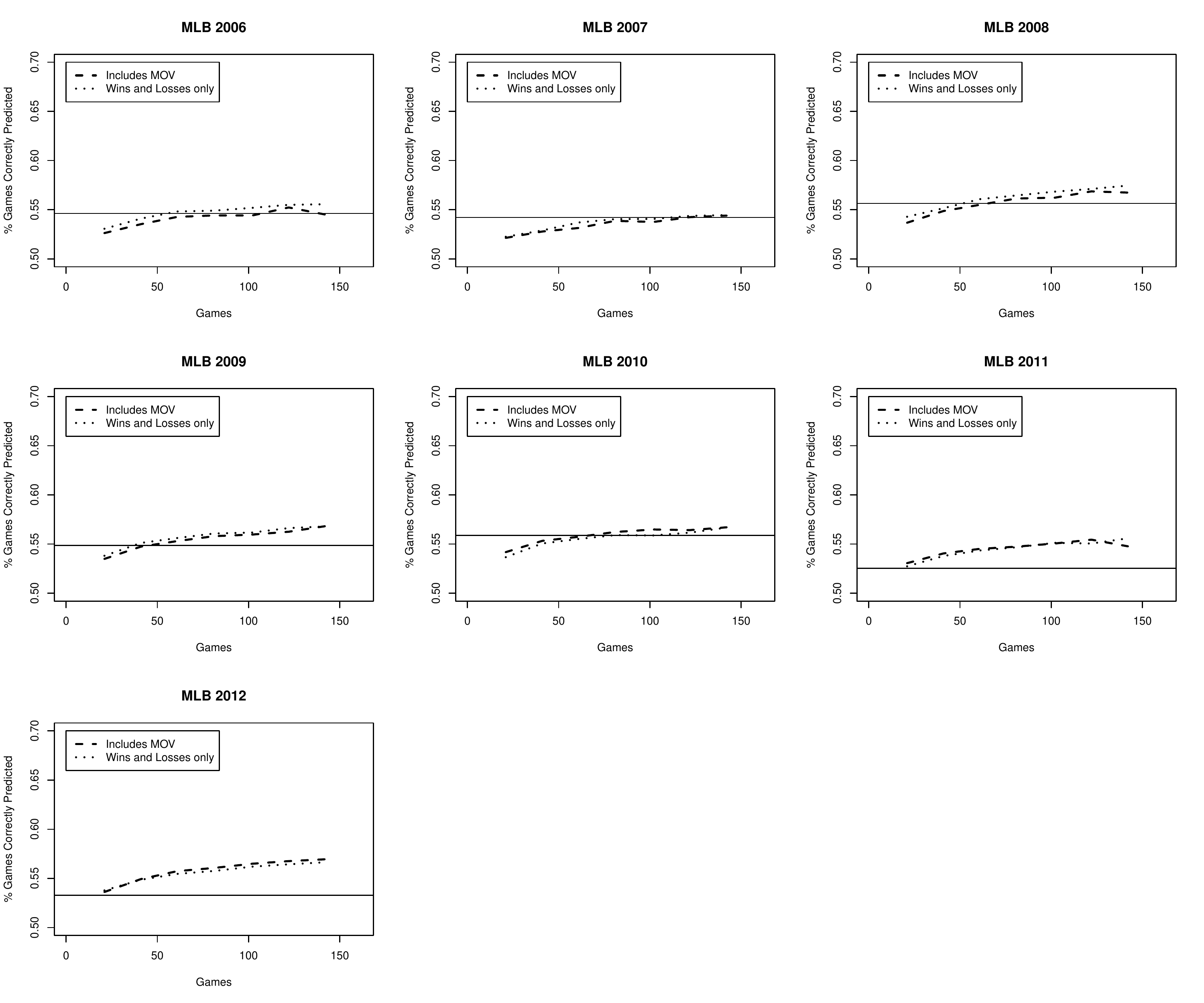}
\caption{Percent of games correctly predicted on test set vs. average number of games per team in training set, MLB seasons 2006-2012. \label{mlb_results}}
\end{figure}
\begin{figure}[!h]
\centering
\includegraphics[width = 6in]{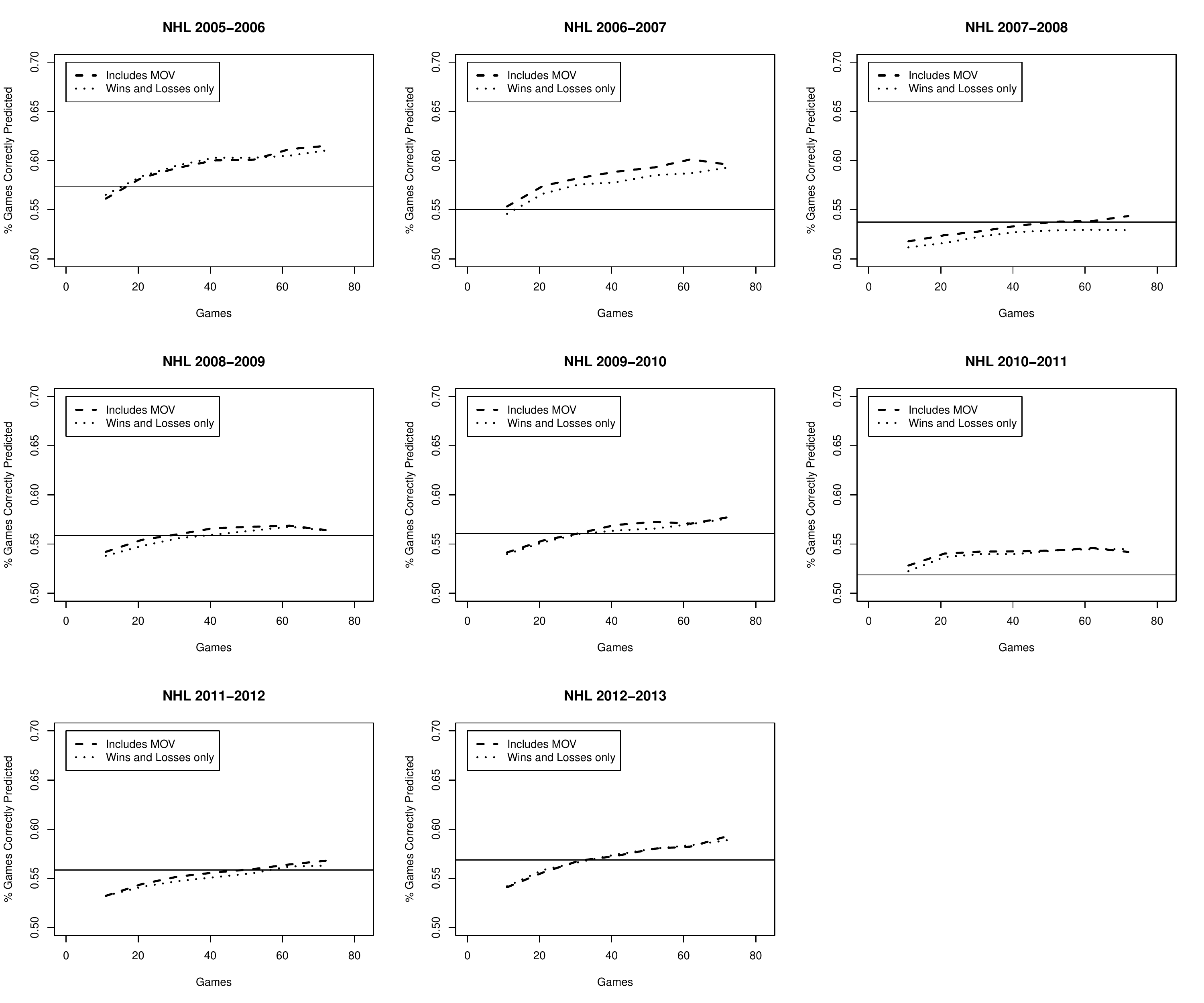}
\caption{Percent of games correctly predicted on test set vs. average number of games per team in training set, NHL seasons 2005-2013. \label{nhl_results}}
\end{figure}
Results from Major League Baseball and the National Hockey League are found in Figures~\ref{mlb_results} and ~\ref{nhl_results}, respectively. The results for MLB and the NHL were quite similar, in that both leagues were substantially less predictable than the NFL and NBA. The percentage of games correctly predicted for MLB never exceeded 58\% even when 140 games (7/8 of a season) were included in the training set. The NHL was slightly better but our model was never able to predict more than 60\% of games correctly (and this was only achieved in the 2005-2006 season when the home team win probability was relatively high at 58\%). More importantly, prediction accuracy was rarely more than 2-3 percentage points better than the simple strategy of picking the home team in every game for either league. In fact, during the 2007-2008 and 2011-2012 seasons picking the home team performed better than paired comparison models constructed using a half-season's worth of game results. 

It is perhaps not surprising that the outcome of a randomly chosen baseball game is hard to predict based on previous game results given the significant role that the starting pitcher plays in determining the likelihood of winning. In a sense, the ``effective season length'' of MLB is far less than 162 games because each team-pitcher pair carries a different win probability. In additional analyses (results not shown), we fit paired comparison models including a starting pitcher effect, but this did not substantially affect our results.

\subsection{Comparing the sports}

\begin{figure}[!h]
\centering
\includegraphics[width=\textwidth]{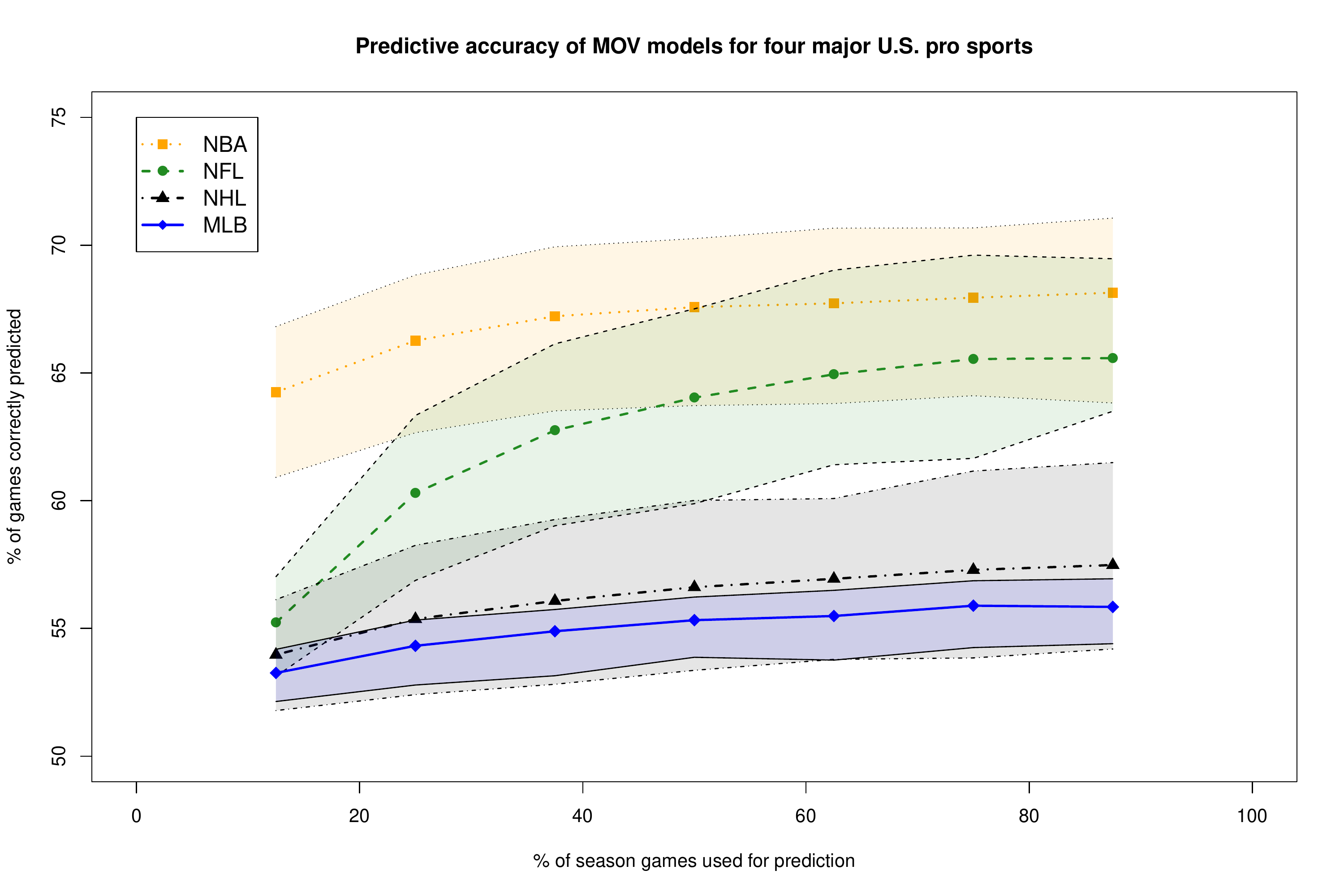}
\caption{Percent of games correctly predicted by margin of victory model on test set vs. percent of season in training set, for four major U.S. sports leagues. For each sport, connected plotting symbols represent the average predictive accuracy and shaded regions enclose the range of predictive accuracies across the seasons of available data. \label{allsports}}
\end{figure}

Figure \ref{allsports} displays curves of summarizing predictive accuracy of the MOV model for the four major sports, aggregated across the years of available data (results from the win-loss model were similar). We see that, even after only 1/8th of the games in a season have been played, substantial information on relative team strength has already accrued in the NBA, while much less can be said at this point about the NFL, NHL, and MLB. Predictive accuracy increases most rapidly with additional games in the NFL, so that predictive accuracy approaches that of the NBA when a substantial fraction of games are used for prediction. As seen above, the overall predictive accuracies for the MLB and NHL remain low, and do not increase markedly with the fraction of games in the training set.

Table \ref{EOS.OR} gives one way of summarizing the informativeness of games in the four major sports, via an odds ratio comparing the predictive accuracy of two models: 1) the MOV paired comparison model using game data from 87.5\% of the season, and 2) a prediction ``model'' which always picks the home team. There is a clear separation between the NFL/NBA, where games played during the season improve the odds of making correct predictions by about 40\% over a ``home-field advantage only'' model, and the NHL/MLB, where the improvement is under 10\%.

\begin{table}[!h]
\centering
\caption{Odds ratio comparing the predictive accuracy of a MOV paired comparison model using data from 87.5\% of a season to the accuracy of a model which always picks the home team. \label{EOS.OR}}
\begin{tabular}{l|c}
 & OR\\
 \hline
 \textbf{NBA} & 1.41\\
\textbf{NFL} & 1.46\\
\textbf{NHL} & 1.09\\
\textbf{MLB} & 1.06
\end{tabular}
\end{table}

Table \ref{infopergame} summarizes the per-game rate of increase in predictive model accuracy for the four sports. The estimates are obtained by fitting least-squares regression lines to the data displayed in Figure \ref{allsports}. The lines for each sport are constrained to a an intercept of 0.5, representing the predictive accuracy of a ``no-information'' model before any games have been played. In developing prediction models for actual use, one might want to incorporate prior information on the home-field advantage based on previous seasons, but in our simple paired comparison models both team strengths and the home-field advantage are estimated purely from current-season data. Hence, prior to any games being played these models can perform no better than flipping a fair coin. The columns of Table \ref{infopergame} correspond to the estimated rate of increase in predictive accuracy, on a percentage point per game scale, over 25\%, 37.5\%, 50\% and 87.5\% of the season. 

\begin{table}[!h]
\centering
\caption{Estimated per-game percentage point increase in predictive accuracy of a margin-of-victory model for the four U.S. sports leagues, by percentage games used to train the model. \label{infopergame}}
\begin{tabular}{l|cccc}
 & 25\% of games & 37.5\% of games & 50\% of games & 87.5\% of games\\
\hline
\textbf{NBA} & 0.91 & 0.69 & 0.55 & 0.34\\
\textbf{NFL} & 2.6 & 2.3 & 2 & 1.4\\
\textbf{NHL} & 0.29 & 0.23 & 0.19 & 0.13\\
\textbf{MLB} & 0.12 & 0.094 & 0.079 & 0.053\\
\end{tabular}
\end{table}

The results in Table \ref{infopergame} allow us to compute a ``per-game informativeness ratio'' between pairs of sports. For example, considering the last column allows us to estimate that, over the course the season, NFL games are approximately 4 times more informative than NBA games, which are in turn about 2-3 times more informative than NHL games, which are themselves approximately 2-3 times more informative than MLB games. The ``informativeness ratio'' of NFL to MLB games is on the order of 65, or about 6 times larger than the inverse ratio of their respective season lengths (162/16 $\approx$ 10). In contrast, the ratio comparing NFL to NBA games ($\approx$ 4) is slightly smaller than the inverse ratio of their respective season lengths (82/16 $\approx$ 5). 

\section{Conclusions and discussion}

Our results reveal substantial differences between the major North American sports according to how well one is able to discern  team strengths using game results from a single season. NBA games are most easily predicted, with paired comparison models having good predictive accuracy even early in the season; indeed, since our information metric for the NBA appears to plateau around game 30, an argument could be made that the latter half of the NBA season could be eliminated without substantially affecting the ability to identify the teams most deserving of a playoff spot. NFL game results also give useful information for determining relative team strength. On a per-game basis, NFL contests contain the largest amount of information. With the exception of the 2008 season, there was no obvious ``information plateau'' in the NFL, though the rate of increase in information did appear to slow somewhat after the first 5 games. These results suggest that games in the latter part of the NFL season contribute useful information in determining who the best teams are.

The predictive ability of paired comparison models constructed from MLB and NHL game data remains limited even when results from a large number of games are used. One interpretation of this finding is that, in comparison to the NBA and NFL, games in MLB and the NHL carry little information about relative team strength. Our results may also reflect smaller variance in team strengths (i.e., greater parity) in hockey and baseball: Because our information metric considers the predictive accuracy averaged across all games in the test set, if most games are played between opposing teams of roughly the same strength then most predictive models will fare poorly.  Indeed, the inter-quartile range for winning percentage in these sports is typically on the order of $\sim$20\%, while in football and basketball it is closer to 30\%. Our observation that the hockey and baseball regular seasons do relatively little to distinguish between teams' abilities is reflected in playoff results, where ``upsets'' of top-seeded teams by teams who barely qualified for the postseason happen much more regularly in the NHL and MLB than in the NFL and NBA. One possible extension of this work would be to quantify this effect more formally.

Indeed, given the relative inability of predictive models to distinguish between MLB teams upon completion of the regular season, a compelling argument could be made for increasing the number of teams that qualify for the MLB playoffs since the current 10-team format is likely to exclude teams of equal or greater ability than ones that make it. Using similar logic, one might also argue that if the goal of the playoffs is to identify the best team (admittedly an oversimplification), then perhaps the NBA playoffs are \emph{overly} inclusive as there is ample information contained in regular season game outcomes to distinguish between the best teams and those that are merely average.

More surprising to us was the enormous discrepancy in the informativeness of game results between hockey and basketball, which both currently play seasons of the same length but perhaps ought not to. One possible explanation for why basketball game results more reliably reflect team strength is that a large number of baskets are scored, and the Law of Large Numbers dictates that each team approaches their ``true'' ability level more closely. In contrast, NHL games are typically low-scoring affairs, further compounded by the fact that a large fraction of goals are scored on broken plays and deflections which seem to be strongly influenced by chance. We have not analyzed data from soccer, but it would be interesting to explore whether the ``uninformativeness'' of hockey and baseball game results extends to other low-scoring sports.

Our analysis has several limitations. First, we chose to quantify information via the predictive accuracy of simple paired comparison models. It is possible that using more sophisticated models for prediction might change our conclusions, though we doubt it would erase the sizable between-sport differences that we observed. Indeed, as mentioned above, accounting for starting pitcher effects in our MLB prediction model did not substantially affect the results. Second, it could be argued that team win probabilities change over the course of a season due to roster turnover, injuries, and other effects. By randomly assigning games to our training and test set without regard to their temporal ordering, we are implicitly estimating ``average'' team strengths over  the season, and applying these to predict the outcome of an ``average'' game. We chose a random sampling approach over one which would simply split the season because we wanted to eliminate time trends in team strengths when describing how information accrued as more game results were observed. While our approach does not directly describe how predictive accuracy improves as games are played in their scheduled order, we anticipate that the patterns would be similar to what we observed. 

\bibliographystyle{plainnat}
\begingroup
\sloppy
\bibliography{InfoInGames_arxiv}
\endgroup
\end{document}